# Low-Temperature Conductivity of Weakly Interacting Quantum Spin Hall Edges in Strained-Layer InAs/GaInSb

Tingxin Li[1*], Pengjie Wang[2], Gerard Sullivan[3], Xi Lin[2], Rui-Rui Du[1,2,]

[1]*Department of Physics and Astronomy, Rice University, Houston, Texas 77251-1892, USA*
[2]*International Center for Quantum Materials, School of Physics, Peking University, Beijing 100871, China*
[3]*Teledyne Scientific and Imaging, Thousand Oaks, California 91603, USA*

* tl51@rice.edu

### *Abstract*

We report low-temperature transport measurements in strained InAs/Ga$_{0.68}$In$_{0.32}$Sb quantum wells, which supports time-reversal symmetry-protected helical edge states. The temperature and bias voltage dependence of the helical edge conductance for devices of various sizes are consistent with the theoretical expectation of a weakly interacting helical edge state. Moreover, we found that the magnetoresistance of the helical edge states is related to the edge interaction effect and the disorder strength.

*Introduction*

In the past decade, topological materials have attracted considerable attention due to their peculiar properties. Among them, the quantum spin Hall insulator (QSHI) [1,2] offers a unique platform to study the "helical" one-dimensional (1D) electron system. From a single-particle point of view, electrons in the helical edge state are counter-propagating and spin-momentum locking modes that are protected by time-reversal symmetry (TRS). However, in InAs/GaSb quantum wells (QWs), because of the fact that a relatively small bulk hybridization gap $\Delta$ (~4 meV) opens at a nonzero wave vector $k_{\text{cross}}$ [3-5], the edge Fermi velocity $v_F \sim \Delta/k_{\text{cross}}$ could be very small (in the range of ~ $2 \times 10^4$ ms$^{-1}$ to ~ $5 \times 10^4$ ms$^{-1}$), driving the helical edge electrons into the strongly interacting regime, described as helical Luttinger liquids [5,6]. In an InAs/GaSb QW of Luttinger parameter $K < 1/4$, the helical edge states show Luttinger-liquid behavior, namely, that the measured edge conductance is suppressed at low temperature and low bias voltage as a power law [5].

By strain engineering, the $\Delta$ can be enhanced up to ~20 meV in the InAs/GaInSb QWs [7,8]. Therefore, the value of $K$ can be tuned close to 1/2. Unlike the ordinary Luttinger liquid, such as in semiconductor nanowires [9] or carbon nanotubes [10,11], where arbitrary weak interactions could modify the system properties in a fundamental way, the 1D helical liquid is in principle insensitive to weak interactions due to topological protections [6,12]. From the renormalization group view, the physical properties of helical liquids are divided by several fixed points of $K$ [6]. Therefore, one would expect that the transport properties of the helical edge conductance in strained InAs/GaInSb QWs of $K \sim 1/2$ (weak interaction) should be quite different from the InAs/GaSb QWs of $K \sim 1/4$ (strong interaction). Although the TRS-protected QSHI has already been observed in strained InAs/GaInSb QWs [7], a systematic transport study aiming at the weakly interacting helical edge state is still lacking. In addition, in HgTe QWs, where the helical edge state is also supposed to be weakly interacting [13], there appear to be discrepancies among different experimental studies concerning the temperature and the magnetic field dependence of the helical edge conductance [14-17].

In this paper, we systematically investigate the low-temperature transport properties of weakly interacting helical edge states in strained InAs/Ga$_{0.68}$In$_{0.32}$Sb QWs ($K \sim 1/2$). We find that the helical edge conductance is independent of temperature and bias voltage within a certain range when the edge length is shorter than the edge coherence length $l_\varphi$, and it weakly depends on temperature and bias voltage for the device edge length longer than the $l_\varphi$. Moreover, the response of the helical edge conductance to external magnetic fields is related to the interaction and disorder strength of the helical edge state.

***Bulk transport properties of inverted InAs/Ga$_{0.68}$In$_{0.32}$Sb QWs***

As shown in Fig. 2(g) of Ref. 7, the bulk hybridization gap Δ obtained from the temperature dependent measurements of a Corbino device made by the InAs/Ga$_{0.68}$In$_{0.32}$Sb QWs is about 20 meV (250 K). By performing magneto-transport measurements, the electron density $n$ can be deduced from the Shubnikov-de Haas (SdH) oscillations [18]. We estimate that the $n_{cross}$ value is $\sim 2\times10^{11}$ cm$^{-2}$ by linear fitting the electron density data points (assuming the parallel-plate capacitor model) and extrapolating it to the peak position of the longitudinal resistance $R_{xx}$, as shown in Fig. 1(a). Thus the bands are in a modestly deep inverted regime. According to the formula reported in Ref. [13,19], the estimated $K$ value of the helical edge states in InAs/Ga$_{0.68}$In$_{0.32}$Sb QWs is about 0.43, namely in the weakly interacting regime.

Corbino devices are widely used for studying the bulk state of QSHIs [4,20], since only the bulk conductance contributes to the signal under Corbino geometry. Figure 1(b) shows the conductance per square $G_\square$ versus the front gate voltage $V_{front}$ of a Corbino device made by InAs/Ga$_{0.68}$In$_{0.32}$Sb QWs at different temperature $T$. As the Fermi level is tuned into the bulk gap by gates, the $G_\square$ has been strongly suppressed, indicating that the bulk of the QSH state is quite insulating. Specifically, at 30 mK, the square resistance in the bulk gap is larger than 100 MΩ although it decreases with the increasing of $T$. The bulk resistance per square of the QSHI at $T = 1$ K is still as large as $\sim 1$ MΩ. Figure 1(c) shows the $G_\square$ - $V_{front}$ traces of the same device measured with different bias voltage $V$ at $T \sim 30$ mK. It can be seen that the bulk of the QSHI in InAs/Ga$_{0.68}$In$_{0.32}$Sb QWs also become less insulating under large $V$, presumably due to heating effects.

In-plane magnetic fields $B_{//}$ could quench the bulk hybridization gap due to a relative momentum shifting between the electron and hole band, and it has already been experimentally verified in modestly deep inverted ($n_{cross} > 2\times10^{11}$ cm$^{-2}$) InAs/GaSb QWs [21] and InAs/GaInSb QWs [7]. For a fixed band inversion, the suppression of bulk hybridization gap depends on both the gap size and the width $d$ of the QWs. For a larger $\Delta$, the suppression field is higher; and for a narrower QW, the external magnetic-field induced momentum shifting is smaller, thus the suppression field is also higher. Therefore, for the InAs/Ga$_{0.68}$In$_{0.32}$Sb QWs we used for experiments with a relatively large $\Delta$ and a relatively small $d$, the quenching of the hybridization gap is shown only when $B_{//}$ is above 10 T (Fig. 2(f) in Ref. 7). On the other hand, no-change of bulk insulating characteristics up to $B_{//}$ = 9 T (Fig. 1(d)) is convenient to detect the magnetic field dependence of the helical edge transport without the disturbance of bulk conductance. Under perpendicular magnetic field $B_\perp$, similar to previous studies [4,5,7], the bulk gap become more insulating due to localization effects (data not shown). Overall, the bulk of the QSH state formed in InAs/Ga$_{0.68}$In$_{0.32}$Sb QWs is sufficiently insulating for measuring edge transport within a relatively large range of $T$, $V$, and $B$. In the rest of this paper, we will focus on the transport properties of its helical edge states.

*Temperature and bias voltage dependence of the helical edge conductance*

The bias current dependence and the temperature dependence of $R_{xx}$ - $V_{front}$ traces for a 12 μm × 4 μm Hall bar are shown in Figs. 2(a) and 2(b). The resistance peak measured with 0.1 nA bias current at $T \sim$ 30 mK is about 58 kΩ, which is larger than the quantized value $h/2e^2$, indicating certain backscattering processes occurring in the helical edge. Although the single-particle elastic backscattering process is prohibited in a TRS-protected helical liquid, inelastic backscattering processes [19,22-29] could still happen. An edge coherence length $l_\varphi$ is usually defined as $l_\varphi = L \cdot R_{quantum}/R_{edge}$ [4,7,14], where $L$ is the edge length; $R_{quantum}$ is the quantized resistance of ballistic transport, and $R_{edge}$ is the measured helical edge resistance. For example, here $R_{edge}$ = 58 kΩ of the 12 μm × 4 μm Hall bar corresponds to a $l_\varphi$ of about 2.7 μm. Compared to the strongly interacting edge state in InAs/GaSb QWs [5], much weaker bias voltage dependence and temperature dependence of the $R_{edge}$ have been observed here for the weakly interacting edge state in InAs/Ga$_{0.68}$In$_{0.32}$Sb QWs. Moreover, unlike the strongly interacting case, the helical edge conductance cannot be fitted by a power law as a function of *T*. Instead a logarithmic function fits the data points better, as shown in the inset of Fig. 2(b).

When the edge length is shorter than the $l_\varphi$, electrons transport inside helical edges should be ballistic without any backscattering. Previous experiments have already demonstrated an approximately quantized edge conductance for QSHIs in HgTe QWs [14], InAs/GaSb QWs [4,5,30], and InAs/GaInSb QWs [7]. It is worth pointing out that in the strongly interacting regime, the quantized conductance appears only in the limit where the *eV* or $k_BT$ ($k_B$ is the Boltzmann constant) dominates over the internal interaction energy [5]. Here in an InAs/Ga$_{0.68}$In$_{0.32}$Sb device with single-edge transport configuration [the schematic figure shown in Fig. 2(c)] [30] of edge length ~1.2 μm, quantized resistance of $h/e^2$ has been observed. Remarkably, the quantized plateau persists from 30 mK to 0.85 K with 0.2 nA bias current, and it persists from 0.1 nA to 10 nA at *T* ~ 30 mK (Fig. 2(c)), this behavior is fundamentally different from the Luttinger-liquid behaviors observed in the strongly interacting InAs/GaSb QWs [5]. The $R_{xx}$ plateau decreases at higher *T* and/or *V*, presumably due to an increasing bulk conductivity. Similar results also have been observed in another two-terminal device of 5 μm edge length (mesa width 3 μm), as shown in Fig. 2(d).

The above experimental observations are qualitatively consistent with the theoretical expectation of a weakly interacting helical edge state. In a TRS-protected helical liquid, only the inelastic backscattering is allowed, and inelastic backscattering processes are usually influenced by temperatures. Therefore, for devices with an edge length that is longer than the $l_\varphi$, the helical edge conductance should have some kinds of *T*-dependence, depending on the interaction strength and/or the specific type of backscattering processes [19,22-29]. Note that most theoretical models predict a strong temperature dependence of the helical edge conductivity, which are inconsistent with our observations. For example, an impurity induced two-particle inelastic scattering leads to a $T^{4K}$ ($K > 1/2$) [22,23] temperature-dependent reduction of the edge conductivity in the weakly interacting limit. As for the single particle inelastic scattering, theoretical models predict $T^{2K+2}$ ($K > 2/3$) [22] or stronger [27,28] temperature-dependent reduction of the edge conductivity. The presence of charge puddles due to electrostatic potential fluctuations also could induce the backscattering in helical edge states [25,26]. It is worth mentioning that charge puddles containing an odd number of electrons can act as magnetic impurities, leading to a much weaker temperature dependence above the Kondo temperature in the weakly interacting limit, $R_{edge} \sim A-B*\ln(C/T)$ (A, B, and C are constant) [19] or $R_{edge} \sim \ln^2(T)$ [26]. Although such sub-power-law relations still cannot fit the experimental data shown in Fig. 2(b) very well, they may provide a qualitative explanation for our observations. On the other hand, the helical liquid has a topological stability that is robust to nonmagnetic disorder and weak interaction effects. Thus the quantized edge conductance could be independent of temperature and bias voltage within a certain range when the sample edge length is shorter than the $l_\varphi$.

*Response of edge conductance to external magnetic fields*

When breaking the TRS by applying a magnetic field, more scattering processes could occur in the helical edge, and indeed, we observed that the helical edge conductance decreases under magnetic fields for all measured devices. Specifically, Figs. 3(a) and 3(b) show the $R_{xx}$-$V_{front}$ traces of the 12 μm × 4 μm Hall bar and the 1.2 μm single edge device at $B_\perp = 1$ T and $B_{//} = 1$ T, as compared to the case of zero magnetic field. It can be seen that the edge conductance drops more rapidly under $B_\perp$ due to orbital effects. In addition, Fig. 3(c) shows the differential $R_{edge}$ versus dc bias voltage $V_{dc}$ of the 12 μm × 4 μm Hall bar at $B = 0$ T, $B_\perp = 1$ T, and $B_{//} = 2$ T. Obviously, the differential $R_{edge}$ shows stronger dependence of $V_{dc}$ under magnetic fields, also the external magnetic fields induced increment of the helical edge resistance decreases with the increasing $V_{dc}$. Similar behaviors also have been observed in the *T*-dependent measurements of the $R_{edge}$, as shown in Fig. 3(d).

It is interesting to compare the response of the helical edge conductance to external magnetic fields between the strongly interacting helical edge states and the weakly interacting helical edge states. Experimentally, we choose devices with negligible gate hysteresis, and hold the gate voltage at their $R_{xx}$ peaks, then sweep the magnetic field. For the strongly interacting regime [Fig. 4(a), devices made by the wafer A mentioned in Ref. 5], the $R_{edge}$ shows strong bias voltage dependence, but does not respond to external magnetic fields. On the other hand, for the weakly interacting regime [Figs. 4(b) and 4(c), devices made by InAs/Ga$_{0.68}$In$_{0.32}$Sb QWs], the $R_{edge}$ shows much weaker bias voltage dependence at zero magnetic field, but becomes stronger under magnetic fields.

These results may imply that in the strongly interacting 1D helical liquid, the TRS is spontaneously broken [6,31], so the measured conductance is independent of external magnetic fields. As for the weakly interacting regime, under the circumstance of broken TRS, the 1D helical liquid could be viewed as a spinless 1D quantum wire [32]. As reported in previous studies [9], in a weakly disordered 1D quantum wire, even weak electron-electron interactions could induce significant backscattering processes, and the measured conductance decreases with decreasing $T$ and $V$. In other words, in the weakly interacting regime, the ordinary 1D liquid is more sensitive to disorders and interactions due to the lack of topological protections. As a result, the measured conductance shows stronger $T$ and $V$ dependence than 1D helical liquids under the same level of interaction and disorder strength. On the other hand, external magnetic fields may open a gap in the helical edge state, and the gap could be smeared at larger $V$ or higher $T$, thus the edge conductance shows smaller responses to external magnetic fields at larger bias voltages and/or higher temperatures.

We further examine the edge conductance of InAs/Ga$_{0.68}$In$_{0.32}$Sb QWs under larger in-plane magnetic fields. In a purely in-plane magnetic field, the orbital effect perpendicular to the plane is absent. Notably, the $R_{edge}$ tend to saturate under larger magnetic fields, as shown in Fig. 4(c). An early theoretical calculation [32] which considers the combined effect of external magnetic fields and nonmagnetic disorders shows that the edge conductance could be fully suppressed under a small magnetic field for disorder strength on the order of the bulk energy gap, due to Anderson localizations. However, more recent calculations [33,34] show that the edge conductance could partially survive for moderate magnetic field and disorder strength. Figure 4(d) summarizes the normalized edge magnetoresistance of several samples with different $L$ and $l_\varphi$. It can be seen that for samples with larger $L$ and smaller $l_\varphi$, i.e. the stronger disorder strength, the response of $R_{edge}$ to external magnetic fields is stronger, which is consistent with the theoretical predictions [32,34].

*Conclusion*

In summary, we have studied the low-temperature transport properties of strained InAs/Ga$_{0.68}$In$_{0.32}$Sb QWs, where the helical edge state is weakly interacting ($K \sim 0.43$). Our results indicate that although the electron-electron interaction still exists in helical edge states, it becomes less relevant, thus the system follows the behaviors of a TRS-protected QSHI based on the single-particle picture. This experiment provides a clear comparison of the physical properties between the strongly interacting and the weakly interacting helical liquid. The strained-layer InAs/Ga$_{1-x}$In$_x$Sb QWs will provide a tunable material system for future studies of the distinct Kondo behavior [19] and other exotic phenomena related to weakly interacting helical liquids.

*Acknowledgments* We thank L. I. Glazman and C. X. Liu for helpful discussions. Work at Rice University was funded by NSF Grant No. DMR-1508644 and Welch Foundation Grant No. C-1682; work at Peking University was funded by NSFC Grant No. 11674009 and National Basic Research Program of China (NBRPC) Grant No. 2015CB921101. T. L. gratefully acknowledges Smalley-Curl Postdoctoral Fellowship sponsored by RCQM, Rice University.

**Figure Captions**

**FIG. 1. Transport results of bulk states.** (a) The solid line shows the $R_{xx}$-$V_{back}$ trace of a 75 μm × 25 μm Hall bar made by the InAs/Ga$_{0.68}$In$_{0.32}$Sb QWs. Squares are the electron densities obtained from SdH oscillations, and the dashed line is the linear fitting of the electron densities. (b), (c), and (d) show the temperature dependence, bias voltage dependence, and in-plane magnetic field dependence of a Corbino device (inner/outer diameter 800/1200 μm), respectively. The bias voltage used for (b) and (d) is 0.1 mV. The ac modulation voltage used for (c) is 50 μV.

**FIG. 2. Temperature and bias dependence measurement of the helical edge conductance.** $R_{xx}$-$V_{front}$ traces of the 12 μm × 4 μm Hall bar (a) measured with 0.1 nA, 1 nA, and 10 nA bias current at 30 mK, respectively; (b) measured at 50 mK, 500 mK, and 2 K biased with 0.1 nA current, respectively. The inset of (b) shows the $G_{edge}$ (conductance of the averaged $R_{xx}$ peaks) versus $T$. (c) and (d) illustrate that for the single edge device and the 5 μm length two-terminal device, the quantized $R_{xx}$ plateau is independent of $T$ and $V$ within a certain range. The schematic drawings of the device configuration are also shown in the figure. For the single edge device, two contacts separated by 1.2 μm served as both the current leads and the voltage probes.

**FIG. 3. Response to external magnetic fields.** $R_{xx}$-$V_{front}$ traces of (a) the 12 μm × 4 μm Hall bar and (b) the 1.2 μm single edge device measured with 0.1 nA under $B = 0$ T, $B_{//} = 1$ T, and $B_{\perp} = 1$ T, respectively. (c) Bias voltage dependence of the differential $R_{edge}$ for the 12 μm × 4 μm Hall bar under $B = 0$ T (square), $B_{//} = 2$ T (circle), and $B_{\perp} = 1$ T (triangle) at 30 mK, respectively. The ac modulation current used for measurements is 0.1 nA. (d) Temperature dependence of the $R_{edge}$ for the 12 μm × 4 μm Hall bar under $B = 0$ T (square), and $B_{//} = 2$ T (circle), respectively.

**FIG. 4. Magnetoresistance of helical edge states.** (a) $R_{edge}$ of a 30 μm × 10 μm Hall bar made by a strongly interacting InAs/GaSb QWs as a function of magnetic fields, measured with 1 nA and 100 nA excitation current at 300 mK, respectively. (b) $R_{edge}$ of a 30 μm × 10 μm Hall bar made by the InAs/Ga$_{0.68}$In$_{0.32}$Sb QWs as a function of $B_{\perp}$ with 0.5 nA, 5 nA, and 50 nA bias current at 300 mK, respectively. (c) $R_{edge}$ of the 75 μm × 25 μm Hall bar made by the InAs/Ga$_{0.68}$In$_{0.32}$Sb QWs as a function of $B_{//}$ with 0.1 nA and 1 nA bias current at 300 mK. (d)

Normalized edge magnetoresistance of several samples made by the InAs/Ga$_{0.68}$In$_{0.32}$Sb QWs with different $L$ and $l_\varphi$. The aspect ratio (length/width) of all four samples is 3.

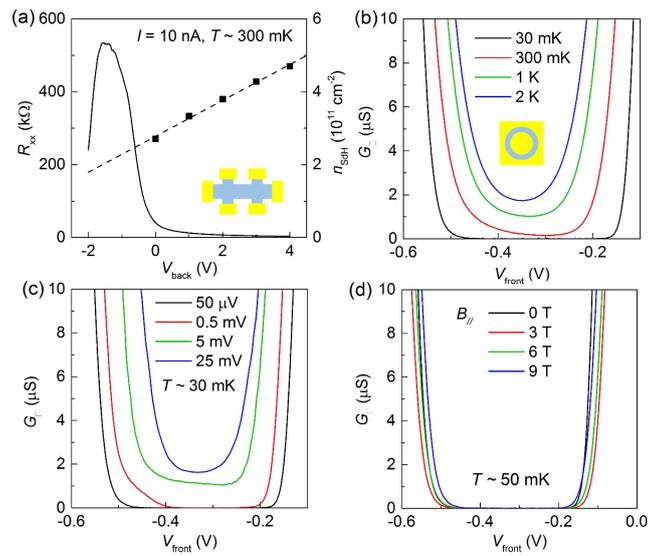

Figure 1

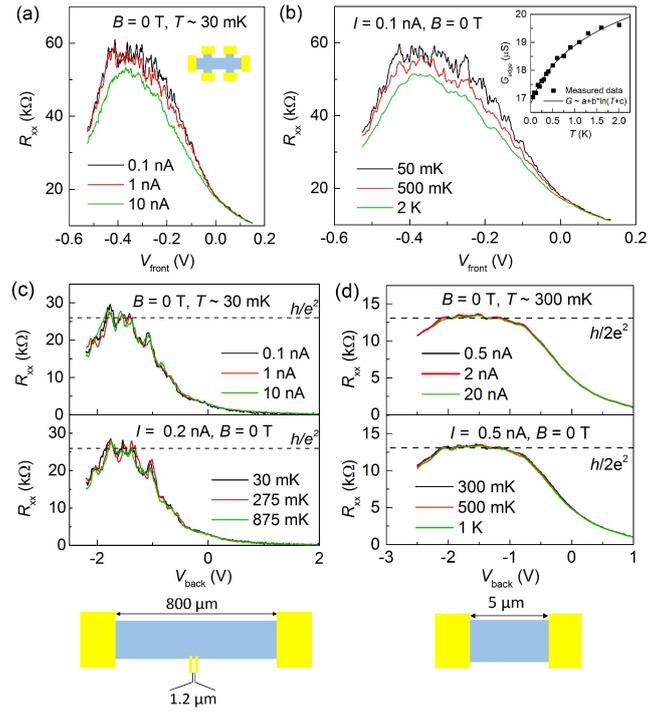

Figure 2

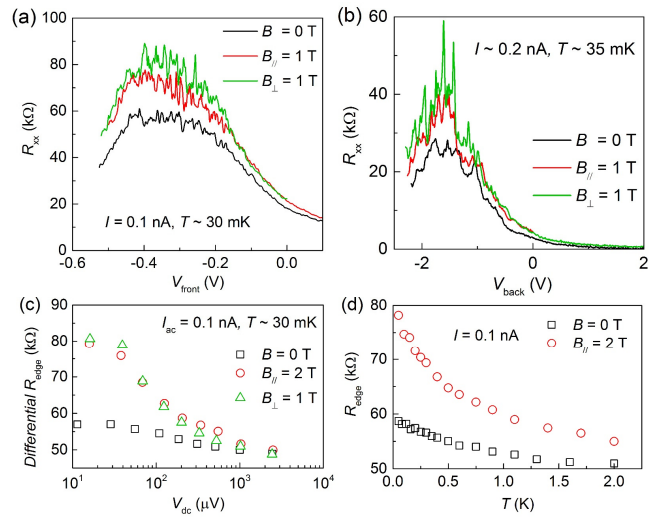

Figure 3

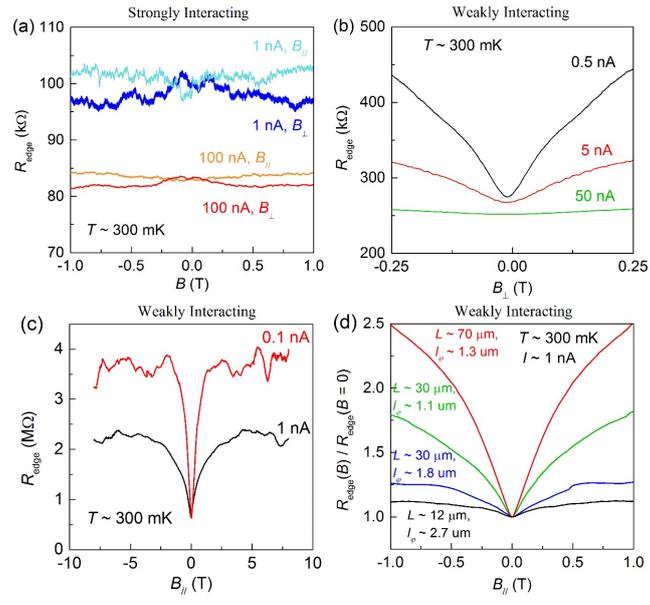

Figure 4

# Supplemental Material

# Low-Temperature Conductivity of Weakly Interacting Quantum Spin Hall Edges in Strained-Layer InAs/GaInSb

Tingxin Li[1*], Pengjie Wang[2], Gerard Sullivan[3], Xi Lin[2], Rui-Rui Du[1,2]

*tl51@rice.edu

[1]*Department of Physics and Astronomy, Rice University, Houston, Texas 77251-1892, USA*
[2]*International Center for Quantum Materials, School of Physics, Peking University, Beijing 100871, China*
[3]*Teledyne Scientific and Imaging, Thousand Oaks, California 91603, USA*

**I Wafer characterizations, device fabrications, and measurement methods**

The semiconductor wafer of 8nm InAs/4nm $Ga_{0.75}In_{0.25}$Sb QWs was grown by MBE technique. Figure S1 is a typical magneto-transport trace of the wafer. Due to the in-plane strain, the mobility of strained InAs/GaInSb QWs are lower than the unstrained InAs/GaSb QWs [S1, S2]. Nevertheless, well resolved Shubnikov-de Haas (SdH) oscillations have been observed for the InAs/$Ga_{0.68}In_{0.32}$Sb QWs, indicating a good quality of the wafer.

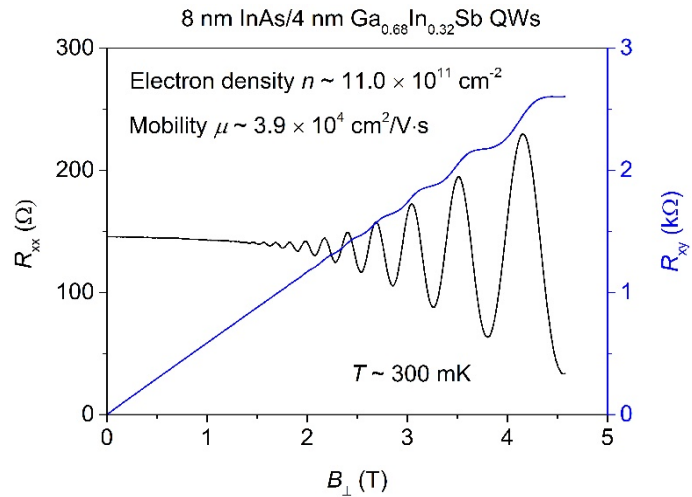

Figure S1: Magneto-transport data of a 75×25 μm$^2$ Hall bar made by the InAs/$Ga_{0.68}In_{0.32}$Sb QWs.

Device processing consisted of the following steps. Mesas were defined by optical and/or E-beam lithography followed by wet etching. Contacts were made by different ways: 1) directly soldering indium at ~ 250 °C; 2) E-beam evaporating palladium (Pd), germanium (Ge), and gold (Au) layers, then annealing at ~ 250 °C in a forming gas ($N_2/H_2$) atmosphere for a few minutes; 3) selectively etching down to the InAs QW, then depositing Ti/Au as electrodes. The device is covered with a thin (30-50 nm) $Al_2O_3$ layer as the front gate dielectric layer, and also for surface passivation. The $Al_2O_3$ layer is grown by atomic layer deposition at ~ 100 °C. Front gates were defined by lithography and then depositing Ti/Au as gate metal.

Low temperature transport measurements were performed in $He^3$ refrigerators of base temperature ~ 300 mK and $He^3$-$He^4$ dilution refrigerators of base temperature ~ 30 mK. A standard low frequency lock-in technique has been used for measurements.

**II Non-local transport data**

The existence of edge states necessarily leads to nonlocal transport. In a non-local measurement configuration (e.g. Fig S2(c)), the voltage probes are far from the bulk current path, so the contribution of bulk conduction to the measured voltage signal is very small. Specifically, the expected Ohm's law contribution to the non-local signal $R_{non-local}$ is ~ $R_{xx} \exp(-\pi L/W)$ [S3], where $R_{xx}$ is the longitudinal resistance measured in the local configuration (Fig S2(a)). For our device, $L$ is 30 μm and $W$ is 10 μm, so the $R_{non-local}$ is suppressed by a factor of ~$10^{-4}$, as compared to the $R_{xx}$. On the other hand, edge transport could lead to a sizeable signal even in the non-local measurement configuration.

Local and non-local measurement data of a of a 30×10 μm$^2$ Hall bar device made by the InAs/$Ga_{0.75}In_{0.25}$Sb QWs are shown in Fig S2(b) and Fig S2(d), respectively. It can be seen that even at 4 K, there exist clear non-local signal when tune the Fermi level into the bulk hybridization gap, where the $R_{non-local}$ is about an order of magnitude smaller as compared to the $R_{xx}$. On the other hand, both in the electron-dominant and hole dominant bulk transport regime, the $R_{non-local}$ is at least three orders of magnitude smaller than the $R_{xx}$. These results illustrate that the edge transport is dominant when the Fermi level is tuned into the bulk gap.

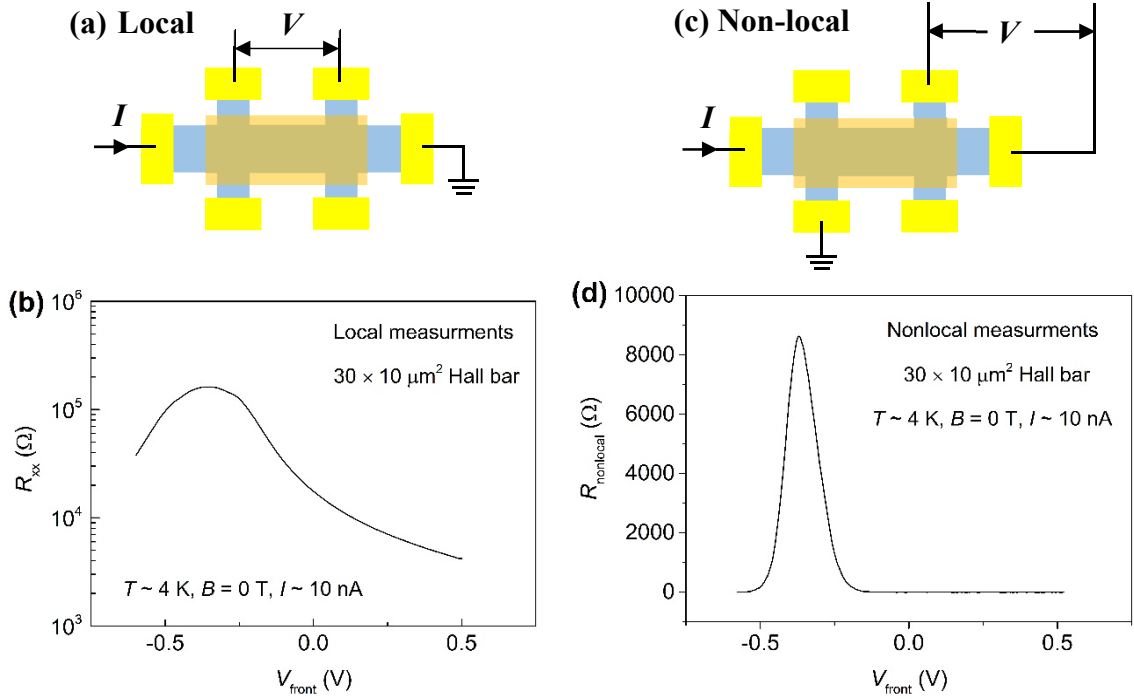

Figure S2: (a) and (c) show the schematic drawings of the electrical configuration for local measurements and non-local measurements, respectively. (b) and (d) show the measured local resistance $R_{xx}$ and non-local resistance $R_{\text{non-local}}$ versus $V_{\text{front}}$, respectively.

**III Transport in non-inverted InAs/GaSb QWs**

Figure S3 shows the $R_{xx}$-$V_{\text{front}}$ traces measured from a 75 μm × 25 μm Hall bar made by an 8 nm InAs/6 nm GaSb QWs. The band structure of this wafer is non-inverted with a normal semiconductor band gap. The devices were fabricated by the same processing methods. It can be seen that the measured resistance in the gap is as large as ~15 MΩ due to the lack of edge conduction, which is consistent with previous studies [S4,S5] of non-inverted InAs/GaSb QWs. This result is fundamentally different from those in the inverted band, proving that there is no 'trivial' edge states [S6,S7] can be observed in our devices.

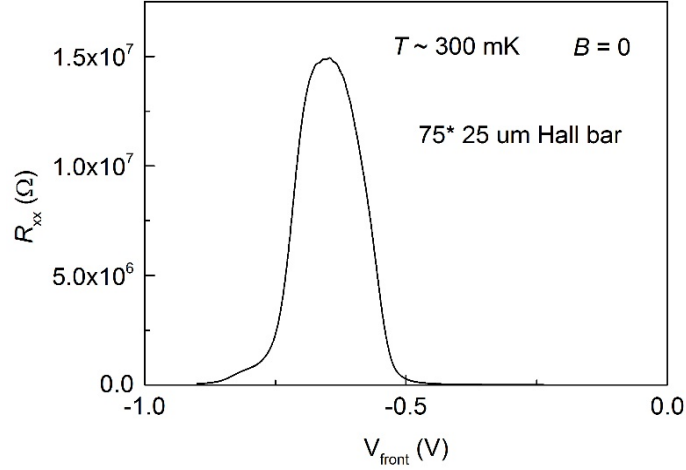

Figure S3: Magneto-transport data of a 75×25 μm² Hall bar made by an 8nm InAs/6nm GaSb QWs.

**Reference**

[S1] I. Knez, R. R. Du, and Gerard Sullivan, Finite conductivity in mesoscopic Hall bars of inverted InAs/GaSb quantum wells. *Phys. Rev. B* **81**, 201301(R) (2010).

[S2] B. M. Nguyen, W. Yi, R. Noah, J. Thorp, and M. Sokolich, High mobility back-gated InAs/GaSb double quantum well grown on GaSb sunstrate. *Appl. Phys. Lett.* **106**, 032107 (2015).

[S3] D. A. Abanin, S. V. Morozov, L. A. Ponomarenko, R. V. Gorbachev, A. S. Mayorov and et al, Giant Nonlocality Near the Dirac Point in Graphene. *Science* **332**, 328 (2011).

[S4] M. J. Yang, C. H. Yang, B. R. Bennett, and B. V. Shanabrook, Evidence of a hybridization gap in "semimetallic" InAs/GaSb Systems. *Phys. Rev. Lett.* **78**, 4613-4616 (1997). In Fig. 4(a) of this paper, the authors showed the transport data of a non-inverted 6 nm InAs/ 6 nm GaSb QWs.

[S5] K. Suzuki, Y. Harada, K. Onomitsu, and K. Muraki, Edge channel transport in the InAs/GaSb topological insulating phase. *Phys. Rev. B* **87**, 235311 (2013). In Fig. 2(a) of this paper, the authors showed the transport data of a non-inverted 10 nm InAs/ 5 nm GaSb QWs.

[S6] F. Nichele, H. J. Suominen, M. Kjaergaard, C. M. Marcus, E. Sajadi, and et al, Edge transport in the trivial phase of InAs/GaSb. *New J. Phys.* **18** 083005 (2016).

[S7] B. M. Nguyen, A. A. Kiselev, R. Noah, W. Yi, F. Qu, and *et al*, Decoupling Edge Versus Bulk Conductance in the Trivial Regime of an InAs/GaSb Double Quantum Well Using Corbino Ring Geometry. *Phys. Rev. Lett.* **117**, 077701 (2016).